\documentclass[aps, pre, reprint, superscriptaddress]{revtex4-1}
\usepackage{bm,epsfig,mathrsfs,amsmath,amssymb, xr}
\usepackage{graphicx}
\usepackage{dcolumn}
\usepackage{color}

\begin{document}

\newcommand\uleip{\affiliation{Institut f\"ur Theoretische Physik, Universit\"at Leipzig,  Postfach 100 920, D-04009 Leipzig, Germany}}
\newcommand\chau{\affiliation{ 
 Charles University,  
 Faculty of Mathematics and Physics, 
 Department of Macromolecular Physics, 
 V Hole{\v s}ovi{\v c}k{\' a}ch 2, 
 CZ-180~00~Praha, Czech Republic 
}}

\title{Underdamped Active Brownian Heat Engine}

\author{Viktor Holubec}\email{viktor.holubec@mff.cuni.cz}\uleip\chau
\author{Rahul Marathe}\email{maratherahul@physics.iitd.ac.in}\affiliation{Department of Physics, Indian Institute of Technology, Delhi, New Delhi 110016, India}

\begin{abstract}
Active Brownian engines rectify energy from reservoirs composed of self-propelling non-equilibrium molecules into work. We consider a class of such engines based on an underdamped Brownian particle trapped in a power-law potential. The energy they transform has thermodynamic properties of heat only if the non-equilibrium reservoir can be assigned a suitable effective temperature consistent with the second law and thus yielding an upper bound on the engine efficiency. The effective temperature exists if the total force exerted on the particle by the bath is not correlated with the particle position. In general, this occurs if the noise autocorrelation function and the friction kernel are proportional as in the fluctuation-dissipation theorem. But even if the proportionality is broken, the effective temperature can be defined in restricted, fine-tuned, parameter regimes, as we demonstrate on a specific example with harmonic potential.
\end{abstract}

\maketitle
\date{\today}

\section{Introduction}
The surging field of active matter has recently attracted a lot of attention of researchers from various fields~\cite{Prost2015,Popkin2016,Needleman2017,Doostmohammadi2018,Feinerman2018,Trepat2018,Xi2019}. It studies the behavior of self-propelling agents ranging from suspensions of micrometer-sized Janus particles and bacteria to flocks of birds~\cite{Sriram2010, Bechinger2016, Das2020}. Besides focusing on dynamics of these systems~\cite{Libchaber2000,Cates2008,Cates2015}, several studies attempted to put the active matter on firm thermodynamic footing~\cite{Maggi2014,Fodor2016, Szamel2019}. The main aim of these efforts, which include investigation of proper definitions of the entropy production and derivation of corresponding fluctuation theorems for active matter~\cite{Mandal2017,Rajarshi2018,Dabelow2019}, is to develop a consistent generalization of successful theoretical framework of stochastic thermodynamics~\cite{Sekimoto1998,Seifert2012} to active matter systems~\cite{Speck2016}. 

One of the most intriguing questions in this respect is when the energy extracted from active baths can be termed as heat~\cite{Holubec2020} -- the problem which never arises for equilibrium heat reservoirs providing only heat. While it is straightforward to identify the extracted energy as work for various ratchets~\cite{Leonardo2010,Reichhardt2017,Pietzonka2019} rectifying the directed active motion of the non-equilibrium constituents of the bath, it is not that simple for cyclically operating machines~\cite{Krishnamurthy2016,Zakine2017,Marathe2018,Cates2018,Marathe2019,Ekeh2020,Holubec2020,Lahiri2020,Lahiri2020n}. With respect to the latter, the extracted energy can be unambiguously termed as heat only if there exists an equivalent setup with an equilibrium bath and the same average energy currents as the active engine~\cite{Holubec2020}. If such an equilibrium mapping exists, the non-equilibrium setups can outperform the equilibrium ones~\cite{Seifert2007, Blickle2012, Arnab2014, Holubec2014, Martinez15, Fodor2016, Edgar2016,Edgar2017, Holubec2017, Holubec2018, Pietzonka2018, Wexler2020}
\begin{figure}[!thbp]
   \includegraphics[height=4.5cm,width=8cm,angle=0]{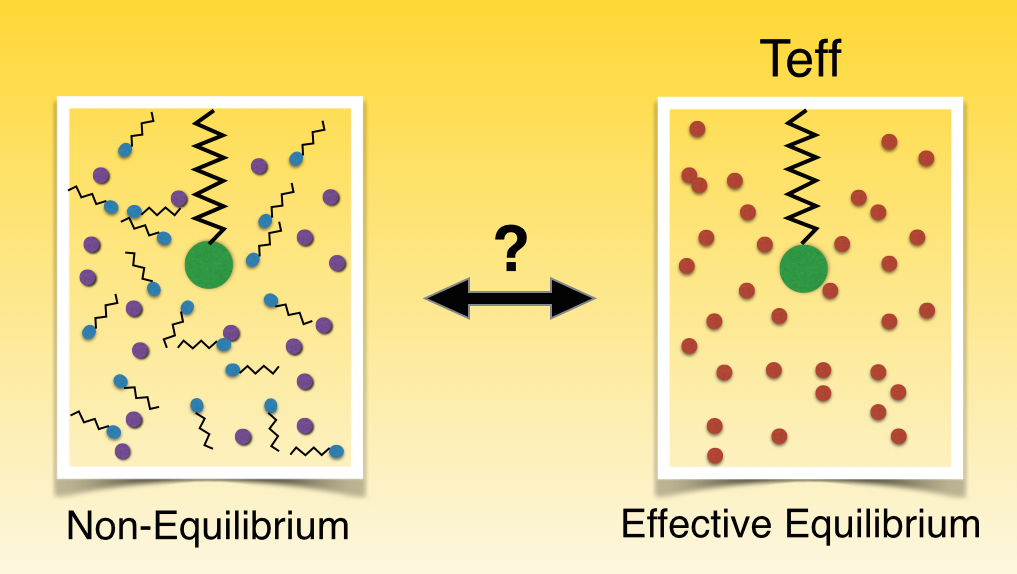}
    \caption{Our setup. Left: An externally controlled working medium of an active Brownian engine (green particle attached to a spring) extracts energy from a non-equilibrium bath. Right: This energy can be termed as heat only if the energy fluxes in the engine can be realized using an equilibrium bath at an effective temperature $T_{\rm eff}$.}
    \label{cartoon}
\end{figure}   
only by achieving unnaturally high effective temperature differences (and thus efficiencies), allowed by the lack of thermalization in the active bath. For example, the recently realized engine with a bath containing living bacteria was reported to operate at effective temperatures far beyond the boiling point of water while the background fluid was still at the room temperature~\cite{Krishnamurthy2016}. If such a mapping does not exist, the active engines can easily break the second law limitations, with the most striking example being the cyclic energy extraction from a single bath reported in Ref.~\cite{Ekeh2020}. Such machines should thus be termed as a (lossy) work-to-work converters.

The most general setting where the equilibrium mapping generically exists are heat engines with working medium described by Hamiltonians of the form $\mathcal{H} = k(t) h(\mathbf{x},\mathbf{p})$, where $k(t)$ is an externally controlled parameter (e.g., playing the role of the inverse volume in macroscopic heat engines) and $h$ is an arbitrary function increasing sufficiently fast with the absolute position $|\mathbf{x}|$ and velocity $|\mathbf{v}|$~\cite{Holubec2020}. 

A typical situation considered in this work is depicted in Fig. \ref{cartoon}. We show when the equilibrium mapping exists for cyclic machines with the working medium described by the underdamped Hamiltonian
\begin{equation}
\mathcal{H}(x,v,t) = \mathcal{V}(x,t) + \frac{1}{2} m v^2,
\label{eq:Hamiltonian}
\end{equation}
where $m$ is a constant mass and $\mathcal{V}(x,t) = \frac{1}{n} k(t) x^n$, $n=2,4,\dots$, denotes a confining potential with periodically modulated stiffness $k(t)$. Existence of the equilibrium mapping in this case is generally not guaranteed, and we have to consider an explicit model. Inspired by recent experiments on Brownian heat engines~\cite{Blickle2012,Edgar2016,Krishnamurthy2016}, we assume that the working medium is an underdamped Brownian particle described by the system of Langevin equations
\begin{eqnarray}
\dot{x} &=& v,
\label{eq:dxGen}\\
m\dot{v} &=& -kx^{n-1} + F + \eta.
\label{eq:dvGen}
\end{eqnarray}
The force $- k x^{n-1} = - \partial_x \mathcal{V}$ corresponds to the potential and $F + \eta$ is the total force applied on the particle by the bath. Its systematic component 
\begin{equation}
F \equiv -\int_{-\infty}^t dt'\, \Gamma(t-t') v(t')
\end{equation}
is a friction force with a friction kernel $\Gamma(t)$. And the 
additive noise with zero mean, $\eta$, denotes its stochastic component. We do not assume that the friction and noise fulfill the (second) fluctuation-dissipation (FDT) theorem~\cite{Kubo1966,Kubo2012,Zwanzig2001} 
\begin{equation}
k_{\rm B} T\Gamma(|t|) = \left<\eta(t)\eta(0)\right>
\label{eq:FDTintro}
\end{equation}
and thus the bath can be out of equilibrium. The dynamics~\eqref{eq:dxGen}--\eqref{eq:dvGen} might for example describe an underdamped active Brownian particle and thus, from now on, we term the considered engine as the underdamped active Brownian engine (UABE).

The energy fluxes into the UABE can be identified from the change of the average internal energy of the working medium \cite{Sekimoto1998}
\begin{equation}
\frac{d}{dt}\left<\mathcal{H}\right> = \frac{d}{dt}\left(\frac{1}{n}k\sigma_x + \frac{1}{2}m\sigma_v \right) = \dot{W} + \dot{Q},
\label{eq:1st_law}
\end{equation}
where the average is taken over realizations of the noise $\eta$ and $\sigma_x \equiv \left<x^n \right>$ and $\sigma_v \equiv \left<v^2 \right>$. The energy per unit time flowing into the system due to the external driving,
\begin{equation}
\dot{W} = \frac{1}{n} \dot{k} \sigma_x,
\label{eq:dW}
\end{equation}
is the input work flux. The rest of the total energy influx,
\begin{equation}
\dot{Q} = \frac{d}{dt}\left<\mathcal{H}\right> - \dot{W} =  \frac{1}{n} k \dot \sigma_x +  \frac{1}{2} m  \dot \sigma_v,
\label{eq:dQ}
\end{equation}
originates in the non-equilibrium bath. While we denote it using the standard symbol for heat flux, it has thermodynamic properties of heat described by the second law only if there exists the equilibrium mapping~\cite{Holubec2020}, i.e. if there is an engine with equilibrium bath with the same energy fluxes~\eqref{eq:dW} and \eqref{eq:dQ} as the UABE. In such a case, the ratio of the output work and the input heat,
\begin{equation}
E \equiv \frac{W_{\rm out}}{Q_{\rm in}} \equiv \frac{ - \int_0^{t_{\rm p}} dt\, \dot{W}(t)}{\int_0 ^{t_{\rm p}} dt\, \dot{Q}(t) \theta{\left[\dot{Q}(t)\right]}},
\label{eq:effc}
\end{equation}
measuring the efficiency $E$ of the UABE with period $t_{\rm p}$, is bounded by the standard second-law bound corresponding to the given cycle realized by the stiffness $k(t)$ and the effective temperature $T_{\rm eff}(t)$. The Heaviside Theta function $\theta$ in the definition of the input heat $Q_{\rm in}$ ensures that the integral evaluates only the average heat flowing into the system ($\dot{Q}(t)>0$)~\cite{Holubec2020}.

In this paper, we focus on the upper bounds on thermodynamic efficiency of UABEs and thus we study existence of the equilibrium mapping in the quasi-static limit, where the control parameter $k$ changes on time-scales much longer than the system relaxation time, the frictional losses are minimal, and the total effective entropy $\Delta S_{\rm tot} \equiv -\int_0^{t_{\rm p}} dt \dot{Q}(t)/T_{\rm eff}(t)$ produced per cycle vanishes~\cite{Holubec2020}.  Then the efficiency~\eqref{eq:effc} can be evaluated using the standard equilibrium thermodynamics. Namely, if the effective temperature and the stiffness are changed in such a way that the resulting cycle is composed of two branches with a constant effective temperature and two adiabatic branches when $\dot{Q}(t) = 0$, the efficiency of the cycle is given by the Carnot efficiency $E_{\rm C} = 1 - T_{\rm eff}^-/T_{\rm eff}^+$, where $T_{\rm eff}^{\mp}$ denote the smallest (-) and largest (+) values of $T_{\rm eff}$ during the cycle. For arbitrary different driving, the efficiency is smaller than $E_{\rm C}$. And it is still given by the standard (equilibrium) formula corresponding to the given cycle (such as the Stirling cycle), if the effective temperature is substituted for the real temperature in these formulas. For a more detailed discussion, see Ref.~\cite{Holubec2020}.

For a quasi-static driving, the dynamics of the moments $\sigma_x$ and $\sigma_v$ in all equilibrium models is described by the combination of equipartition $2\left< \mathcal{T} \right> \equiv \sigma_v = 2T_{\rm eff}$ and virial $2\left< \mathcal{T} \right> = \left<x\frac{\partial{\mathcal{V}}}{\partial{x}} \right>$ theorems:
\begin{equation}
T_{\rm eff} = \frac{1}{2} k \sigma_x = \frac{1}{2} \sigma_v,
\label{eq:TeffEquiPar}
\end{equation}
where $T_{\rm eff}$ denotes the temperature of the corresponding equilibrium baths. Above and in the rest of the paper, we use units in which $k_{\rm B}=1$ and $m=1$. If the noise in Eq.~\eqref{eq:dvGen} can be described by an effective temperature allowing the above described standard equilibrium thermodynamic analysis of the engine efficiency, it must be given by Eq.~\eqref{eq:TeffEquiPar}. In the rest of this paper, we study when such an effective temperature exists.

\section{General results}
\label{sec:EQmodel}

Let us now find the general condition which must be fulfilled in the dynamics of the UABE so that the quasi-static equilibrium mapping~\eqref{eq:TeffEquiPar} exists (for an example of an equilibrium mapping for non-quasi-static protocols, we refer to Ref.~\cite{Holubec2020}). Multiplying Eq.~\eqref{eq:dxGen} by $v$, Eq.~\eqref{eq:dvGen} by $x$, taking the averages, and summing the results, we find that the moments $\sigma_x$ and $\sigma_v$ obey the equation
\begin{equation}
\frac{d}{dt} \left<xv\right> =  - k\sigma_x + \sigma_v + \left<(F +\eta) x\right>.
\end{equation}
In the quasi-static regime, we can neglect the time-derivative on the LHS obtaining the virial theorem
\begin{equation}
\left<\mathcal{T} \right> = -\frac{1}{2} \left<F_{\rm tot} x
\right>
\end{equation}
for the whole system with the total force $F_{\rm tot} = -\partial {\mathcal V}/\partial{x} + F + \eta$ applied to the particle.
The requirement \eqref{eq:TeffEquiPar} implies that the equilibrium mapping exists if the contribution from the bath to the virial vanishes, 
\begin{equation}
\left<(F + \eta) x\right> = 0.
\label{eq:EPTGen2}
\end{equation}
This condition is quite reasonable since it means that the force exerted on the particle by the bath and the particle position are uncorrelated, i.e. that the bath is homogeneous. For an active bath, the condition~\eqref{eq:EPTGen2} might be broken if the particle interacts strongly with the bath. For example, baths composed of active particles get polarized close to walls potentially leading to nonzero $\left<(F + \eta) x\right>$.

The condition~\eqref{eq:EPTGen2}
is naturally fulfilled when the noise autocorrelation function and the friction kernel are proportional as in the FDT~\eqref{eq:FDTintro}. Even though this situation might look trivial since it is mathematically equivalent to settings with equilibrium bath, it still represents an important class of non-equilibrium situations if $T$ in~\eqref{eq:FDTintro} would be given by some effective temperature $T_{\rm eff}$. As an example, consider a system connected to two independent equilibrium reservoirs, $i =1,2$, described by friction forces $F_i = - \gamma_i v$ and white noises $\eta_i = \sqrt{2D_i\gamma_i^2}\xi_i$, $\left<\xi_i(t)\xi_j(t')\right> = \delta_{ij} \delta(t-t')$, with $D_i = T_i/\gamma_i$, as dictated by Einstein relation following from the FDT~\eqref{eq:FDTintro}. Then the total friction force would be $F = -(\gamma_1 + \gamma_2)v$ and the total noise $\eta = \sqrt{2}\sqrt{D_1\gamma_1^2 + D_2\gamma_2^2} \xi$, $\left<\xi(t)\xi(t')\right> = \delta(t-t')$, would be described by the friction kernel $\Gamma(t) = (\gamma_1 + \gamma_2)\delta(t)$ and noise autocorrelation function $\left<\eta(t)\eta(t')\right> = 2 (D_1\gamma_1^2 + D_2\gamma_2^2) \delta(t-t')$. Even though the total noise and the friction obey the FDT~\eqref{eq:FDTintro} with the effective temperature $T_{\rm eff} = (D_1\gamma_1^2 + D_2\gamma_2^2)/(\gamma_1 + \gamma_2)$, the system mediate a nonzero heat flux between the two baths and thus is out of equilibrium whenever $T_1 \neq T_2$. 

Besides these equilibrium-like cases, there are also situations when the noise autocorrelation and friction kernel are not proportional but can be fine-tuned in such a way that Eq.~\eqref{eq:EPTGen2} holds and thus the effective temperature can be defined. Also in these special parameter regimes, the effective temperature~\eqref{eq:TeffEquiPar} can be used to derive upper bounds on efficiency of the corresponding machines. And thus they can serve as solid points for checking general solutions to these systems.

\section{Specific models}
\label{sec:models}

To provide explicit analytical results, we now set $n=2$ in the general discussion of the previous section and thus we resort to the harmonic potential $\mathcal{V}(x,t) = \frac{1}{2} k(t) x^2$. In such a case, it was shown in detail in Ref.~\cite{Holubec2020} how to calculate the effective temperature for one of the key toy models of the active matter, the active Brownian particle model~\cite{Cates2012}. We start by showing that in the underdamped version of this model, the effective temperature in general does not exist.

\subsection{Underdamped active Brownian particle}
\label{sec:NEQmodel}
The underdamped active Brownian particle model is described by the Langevin equations
\begin{eqnarray}
\dot{x} &=& v,
\label{eq:dxnEq2}\\
\dot{v} &=& - \gamma v -kx + \eta,
\label{eq:dvnEq2}
\end{eqnarray}
where $\gamma$ is a friction coefficient, and the stationary zero-mean noise $\eta$ is exponentially correlated, i.e.
\begin{equation}
\left<\eta(t)\eta(t')\right> = \frac{1}{2}\  u^2 \exp(-D_{\rm r} |t-t'|).    
\end{equation}
Here $u$ denotes the swimming velocity of the active particle and $1/D_{\rm r}$ is its orientation decorrelation time.  Since the friction kernel is in this case given by $-\gamma\Gamma(t) = -\gamma \delta(t)$, the noise and the friction are clearly not related by the FDT~\eqref{eq:FDTintro}.

The friction force in this case reads $F = - \gamma v$. Inserting it into the condition \eqref{eq:EPTGen2} for the existence of the equilibrium mapping, we find
\begin{equation}
    - \gamma \left<vx\right> + \left<\eta x\right> = 0.
\end{equation}
From Eq.~\eqref{eq:dxnEq2} it follows that $\left<vx\right> = \dot \sigma_x /2 = 0$ and thus the condition for existence of the equilibrium mapping in this case reads $\left<\eta x\right> = 0$.
In order to calculate this correlation, it is advantageous to rewrite the system~\eqref{eq:dxnEq2} and \eqref{eq:dvnEq2} using the matrix notation as
\begin{equation}
\dot{\bf X} = {\mathcal M} {\bf X} + \eta{\bf e}
\label{eq:dXMatrixEQ},
\end{equation}
where ${\bf X} = (v,x)^\intercal$, ${\bf e} = (1,0)^\intercal$ are column vectors and
\begin{equation}
{\mathcal M} =
\left( {\begin{array}{cc}
   - \gamma & -k \\
   1 & 0 \\
  \end{array} } \right).
\label{eq:Meq}
\end{equation}
The long time solution to Eq.~\eqref{eq:dXMatrixEQ} is given by
\begin{equation}
{\mathbf{X}}(t) = \int^t_{-\infty}dt'\, {\mathcal U}(t-t') \eta(t'){\mathbf e}
\label{eq:Xeqt}
\end{equation}
with ${\mathcal U}(t) = \exp\left[{\mathcal M}(t-t')\right]$. Multiplying it by $\eta(t)$ and taking the average, we find
\begin{figure}[!thbp]
    \includegraphics[height=5cm,width=6.7cm,angle=0]{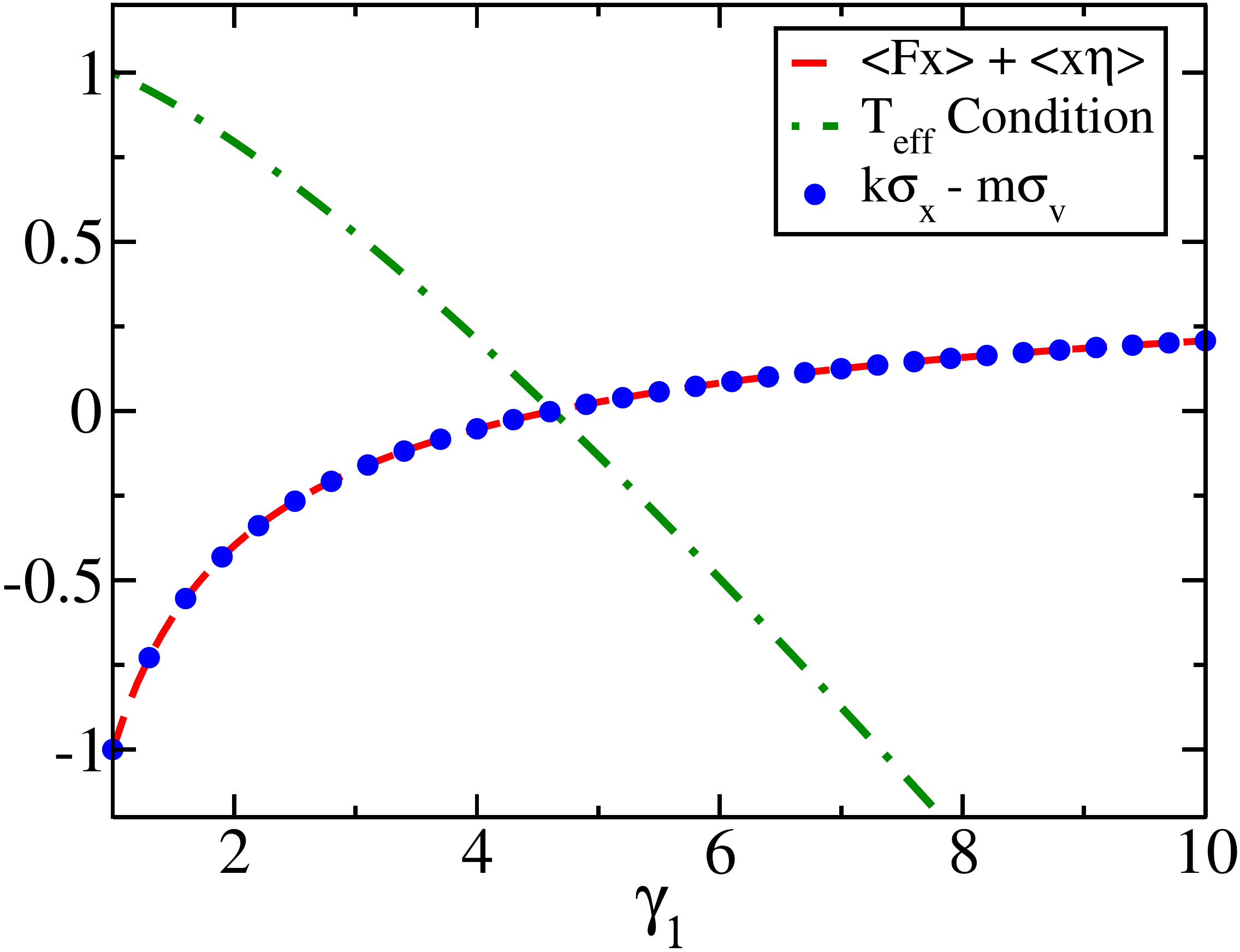}
    \caption{The averages $\langle (F+\eta)x\rangle$ and $k\sigma_x-m\sigma_v$ and the condition~\eqref{eq:condition2} as functions of $\gamma_1$. Parameters used: $\gamma_0 = 4.0$, $\alpha_0 = 0.5$, $\alpha_1 = 1.0$, $m = 1.0$, $k=0.1$ and $D = 1.0$}
    \label{fig1}
\end{figure} 
\begin{multline}
\left<\mathbf{X} \eta \right> = \lim_{t\to \infty}\int^t_0 dt' {\mathcal U}(t-t') \left<\eta(t) \eta(t')\right> {\mathbf e}  = \\
\frac{u^2}{2} \lim_{t\to \infty}\int^t_0 dt' {\rm e}^{{(\mathcal M}-D_{\rm r})(t-t')} {\mathbf e} =  \frac{-1}{{\mathcal M}-D_{\rm r}} \frac{u^2{\mathbf e}}{2},
\label{eq:Xetaneq}
\end{multline}
where the scalar terms are meant to be multiplied by the 2$\times$2 unit matrix. The second element of this vector is the desired correlation between the noise and position. It reads
\begin{equation}
\left<x\eta\right> = \frac{u^2}{2~(k + D_{\rm r} (D_{\rm r} +\gamma))}
\label{eq:xetaneq}
\end{equation}
and it in general does not vanish. The effective temperature for the underdamped Brownian particle thus exists only in various special limiting situations when $\left<x\eta\right> \to 0$. The simplest example is the limit of vanishing swimming velocity of the particle, $u$. In such a case, however, the system dynamics becomes deterministic. Other noteworthy limiting situations where $\left<x\eta\right> \to 0$ are the limits of infinitely strong potential, $k \gg u^2$ and $k \gg D_{\rm r} (D_{\rm r}+\gamma)$, and of infinitely fast orientation decorrelation, $D_{\rm r}^2 \gg k$ and $D_{\rm r}^2 \gg u^2$, when $\eta$ is a white noise.

\subsection{Exponential friction kernel and arbitrary noise}
\label{expCnoisearb}
Let us now consider a slightly more general setup when the friction kernel is exponential, 
\begin{equation}
\Gamma(t) =  \gamma_0\exp(-\gamma_1 |t |)
\label{eq:gammaexp}
\end{equation}
and the noise autocorrelation function is of the form
\begin{equation}
\left<\eta(t)\eta(t') \right> = \alpha_0 \exp(-\alpha_1 |t-t'|) + 2D \delta(t-t').    
\label{eq:expdeltaTCF}
\end{equation}
Also in this case, the proportionality in the FDT~\eqref{eq:FDTintro} is broken. This situation is quite common in experimental setups with active particles suspended in an aqueous medium~\cite{Krishnamurthy2016}, where the working substance experiences random forces from both the active and the solvent molecules. 
 \begin{center}
    \begin{figure}[!thbp]
                \includegraphics[height=5cm,width=7cm,angle=0]{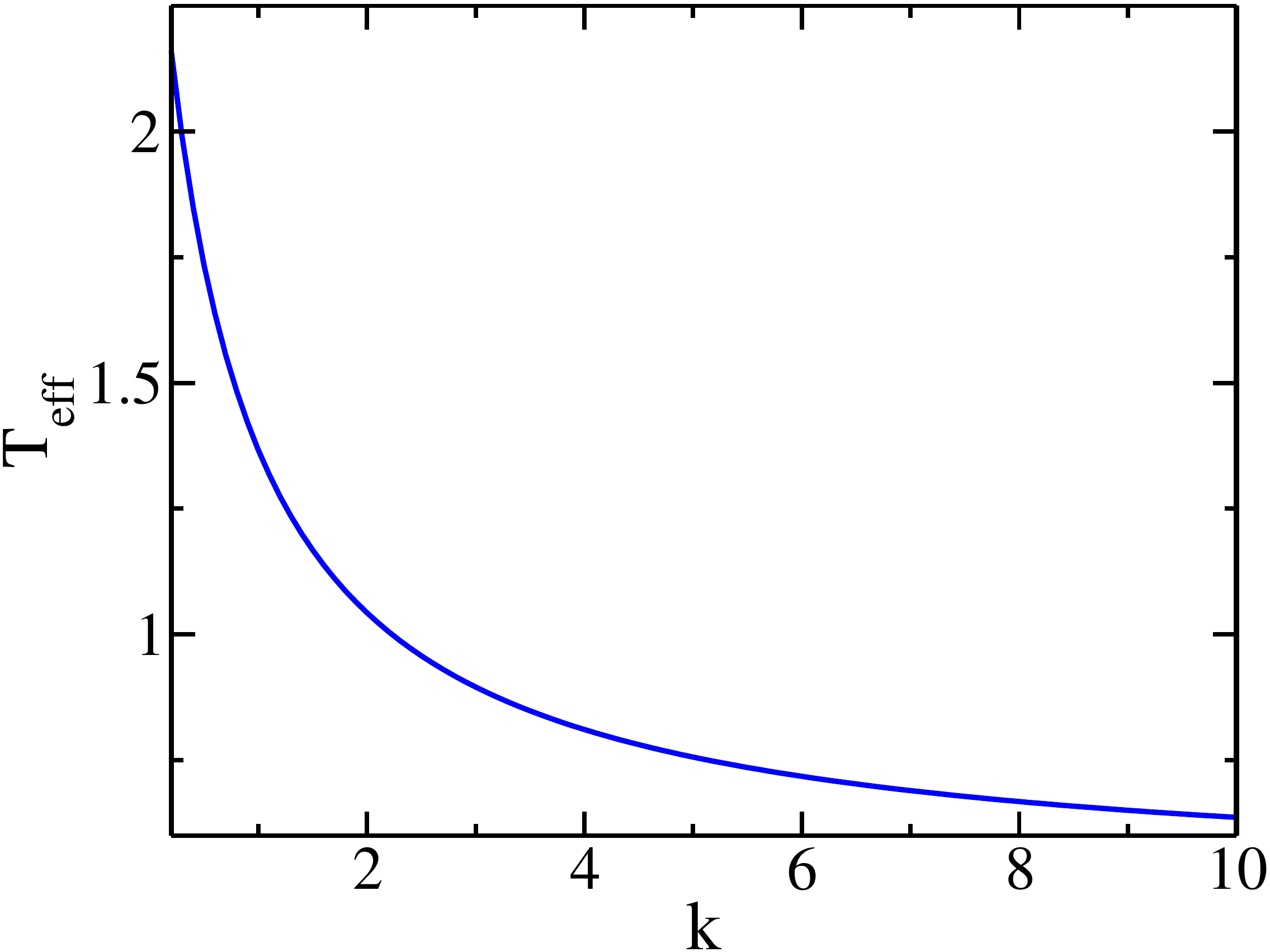}
        \caption{The effective temperature~\eqref{eq:Teff} as a function of the trap strength $k$ for parameters $\gamma_0 = 4.0$, $\gamma_1 = 0.5$, $\alpha_0 = 2.0$, $\alpha_1 = 0.1$, $m = 1.0$ and $D = 1.0$.}
        \label{fig2}
    \end{figure}
\end{center}
Dynamics of this system is described by Eqs.~\eqref{eq:dxGen} and \eqref{eq:dvGen} with the friction kernel~\eqref{eq:gammaexp}. However, for analytical considerations, it is useful to note that the resulting friction force
\begin{equation}
F(t) = -\gamma_0\int_{-\infty}^t dt' \exp\left[-\gamma_1 (t-t')\right]v(t')
\end{equation}
allows one to rewrite the dynamical equations in the form
\begin{eqnarray}
\dot{x} &=& v,
\label{eq:dxEq}\\
\dot{v} &=& -k x + F + \eta,
\label{eq:dvEq}\\
\dot{F} &=& - \gamma_1 F - \gamma_0 v,
\label{eq:dzEq}
\end{eqnarray}
where we assume that all the variables are initialized at $t'=-\infty$.
To find when the condition~\eqref{eq:EPTGen2} for the existence of the equilibrium mapping is fulfilled, we have to calculate the averages $\left<F x\right>$ and $\left<x\eta\right>$. To this end, we again rewrite the system \eqref{eq:dxEq}--\eqref{eq:dzEq} in the matrix form~\eqref{eq:dXMatrixEQ}. This time with
${\bf X} = (F,v,x)^\intercal$, ${\bf e} = (0,1,0)^\intercal$, and
\begin{equation}
{\mathcal M} = \left( {\begin{array}{ccc}
   - \gamma_1 & -\gamma_0 & 0 \\
   1 & 0 & -k \\
	0 & 1 & 0 \\
  \end{array} } \right).
\label{eq:Meq1}
\end{equation}
The long time solution to the resulting matrix equation is again of the form \eqref{eq:Xeqt} with ${\mathcal U}(t) = \exp\left[{\mathcal M}(t-t')\right]$.

To calculate the averages involved in the condition~\eqref{eq:EPTGen2}, we evaluate the whole covariance matrix $\left<{\mathbf{X}}(t) {\mathbf{X}}^\intercal(t)\right>$. Taking its time-derivative derivative, expressing 
$\dot{ \mathbf{X}}$ from Eq.~\eqref{eq:dXMatrixEQ}, and setting $d\left<{\mathbf{X}}(t) {\mathbf{X}}^\intercal(t)\right>/dt = 0$ which follows form the steady-state assumption, we get
\begin{multline}
{\mathcal M} \left<{\mathbf{X}}(t) {\mathbf{X}}^\intercal(t)\right> +
\left<{\mathbf{X}}(t) {\mathbf{X}}^\intercal(t)\right> {\mathcal M}^\intercal \\
+ \left<{\bm \xi} \mathbf{X}^\intercal\right>
+ \left<\mathbf{X} {\bm \xi}^\intercal\right> = 0,
\label{eq:XX}
\end{multline}
where we introduced the shorthand ${\bm \xi}= \eta {\bf e}$. The correlation matrix $\left<{\bm \xi} \mathbf{X}^\intercal\right>^\intercal = \left<\mathbf{X} {\bm \xi}^\intercal\right>$ can be obtained by multiplying the formal solution~\eqref{eq:Xeqt} by ${\bm \xi}$ and taking the average, similarly as in Eq.~\eqref{eq:Xetaneq}. 

The result is~\cite{Caprini2020}
\begin{multline}
\left<\mathbf{X} {\bm \xi}^\intercal\right> =
\lim_{t\rightarrow\infty} \int^t_{-\infty} dt'~ \exp\left[{\mathcal M}(t-t')\right] \left<\eta(t)\eta(t') \right> {\bf e} {\bf e}^\intercal\\
= \left(\frac{\alpha_0}{\alpha_1-{\mathcal M}} + D\right) {\bf e} {\bf e}^\intercal.
\label{eq:Xxi}
\end{multline}
Solution of this linear system is straightforward but the full result is rather lengthy. For our purposes, it is enough to explicitly evaluate only the correlation $\left<Fx\right>$, involved in the condition~\eqref{eq:EPTGen2}, and the variances $\sigma_x$ and $\sigma_v$, defining the effective temperature~\eqref{eq:TeffEquiPar}, if the former condition is fulfilled. Evaluating also the correlation $\left<\eta x\right>$ given by the matrix element 3-2 of the matrix $\left<\mathbf{X} {\bm \xi}^\intercal\right>$ in Eq.~\eqref{eq:Xxi}, we find
\begin{equation}
\left<(F+\eta)x\right> = k\sigma_x-\sigma_v = \frac{(\gamma_1-\alpha_1)(\gamma_1+\alpha_1)c - D}{\gamma_1},
\label{eq:condition1}
\end{equation}
where 
\begin{equation}
c = \frac{\alpha_0}{\gamma_0 \alpha_1 + (\alpha_1 + \gamma_1)~(k+\alpha_1^2)}.
\label{eq:C}
\end{equation}
The equilibrium mapping and the equipartition~\eqref{eq:TeffEquiPar} in this system is thus fulfilled whenever 
\begin{equation}
D  - (\gamma_1-\alpha_1)(\gamma_1+\alpha_1)c = 0,
\label{eq:condition2}
\end{equation}
which can occur even when the FDT is broken. To check validity of our calculations, we also evaluated the terms $\left<(F+\eta)x\right>$ and $k\sigma_x-\sigma_v$ in Eq.~\eqref{eq:condition1} using an independent numerical method. In Fig.~\ref{fig1}, we plot the obtained result and also the explicit condition~\eqref{eq:condition2} as functions of the decay rate $\gamma_1$. The three lines indeed intersect at zero at the same point, proving our analytical calculations. 

Setting, $D = (\gamma_1-\alpha_1)(\gamma_1+\alpha_1)c$, we find that the effective temperature $T_{\rm eff} = k\left<x^2\right>/k_{\rm B}=\left<v^2\right>/k_{\rm B}$ for the present model reads
\begin{equation}
  T_{\rm{\rm eff}} = \frac{\alpha_0}{\gamma_0} \left[ \frac{\gamma_0 \gamma_1 + (\alpha_1 + \gamma_1)~(k+\gamma_1^2 )}{\gamma_0 \alpha_1 + (\alpha_1 + \gamma_1)~(k+\alpha_1^2)}\right].
  \label{eq:Teff}
 \end{equation}
Noteworthy, as shown in Fig.~\ref{fig2}, this effective temperature depends on the trap stiffness $k$. Since similar dependence has been found also for the overdamped setting~\cite{Holubec2020}, this result seems quite general. It is thus puzzling why recent experimental work~\cite{Krishnamurthy2016}, considering a similar setup as ours, reported effective temperature independent of the trap stiffness.

\section{Conclusion}
\label{sec:conclusion}

The existence of mapping to an equivalent equilibrium setup is an important condition for calling machines operating in contact with a non-equilibrium bath as heat engines. Only if such a mapping exists, the efficiency of these machines obeys standard second law limitations. We found that a broad class of machines based on an underdamped Brownian particle trapped in a power law potential, which we called as UABEs, allows for the equilibrium mapping if the total force exerted on the particle by the bath and the particle position are not correlated.

Validity of this condition can be easily checked in experiments. It is always fulfilled if the friction kernel and noise autocorrelation function in the Langevin equation for dynamics of the particle are proportional so that an effective variant of the second fluctuation-dissipation theorem holds. Besides this somewhat trivial case, equilibrium mapping also exists for special parameter regimes in systems where the proportionality is broken. These regimes may serve as solid points where the maximum efficiency of UABEs is known. 

We have studied an explicit example, where such a special parameter regime exists and calculated the effective temperature of the corresponding equilibrium bath. This effective temperature depends on the strength of the applied potential, which, we believe, is a general feature of effective temperatures.

Our findings can guide theoretical analysis and serve as a sanity check of results measured for systems in contact with non-equilibrium reservoirs. As an outlook, it would be interesting to study extensions of our model to finite time regimes, where power delivered by the engines does not vanish. Furthermore, it would be interesting to extend our results to higher dimensions and more complicated potentials.

\begin{acknowledgments}
VH gratefully acknowledges discussions with Klaus Kroy and support by the Humboldt foundation and by the Czech Science Foundation (project No. 20-02955J). RM
acknowledges financial support from Department of Science and Technology (DST), India under the Matrics grant No. MTR/2020/000349.
\end{acknowledgments}




\end{document}